# Fermi gas energetics in low-dimensional metals of spessial geometry

Avto Tavkhelidze*, Vasiko Svanidze, Irakli Noselidze

*Tbilisi State University, Chavchavadze Avenue 13, 0179 Tbilisi, Georgia*

Changes in the metal properties, caused by periodic indents in the metal surface, have been studied within the limit of quantum theory of free electrons. It was shown that due to destructive interference of de Broglie waves, some quantum states inside the low-dimensional metal become quantum mechanically forbidden for free electrons. Wave vector density in k space, reduce dramatically. At the same time, number of free electrons does not change, as metal remains electrically neutral. Because of Pauli exclusion principle some free electrons have to occupy quantum states with higher wave numbers. Fermi vector and Fermi energy of low-dimensional metal increase and consequently its work function decrease. In experiment, magnitude of the effect is limited by the roughness of metal surface. Rough surface causes scattering of the de Broglie waves and compromise their interference. Recent experiments demonstrated reduction of work function in thin metal films, having periodic indents in the surface. Experimental results are in good qualitative agreement with the theory. This effect could exist in any quantum system comprising fermions inside a potential energy box of special geometry.

8530 St

* E-mail: avtotav@geo.net.ge

## Introduction

Recent developments of nanoelectronics enable fabrication of structures with dimensions comparable to the de Broglie wave length of a free electron inside a solid. This new technical capability makes it possible to fabricate some microelectronic devices such as resonant tunneling diodes and transistors, super lattices, quantum wells and others[1] working from the wave properties of the electrons. In this paper we will discuss what happens when regular indents, causing interference of de Broglie waves, are fabricated on the surface of a thin metal film. We will study the free electrons inside a rectangular potential energy box with indented wall and compare results to the case of electron in the box with plane walls. We have shown that modification of the wall of rectangular potential energy box leads to increase of the Fermi energy level. Results obtained for potential energy box were extrapolated to the case of low-dimensional metals (thin metal films). Experimental possibility of fabrication of such indents on the surface of a thin metal film was studied. Practical recommendations regarding dimensions and shape of the indents were given. In addition, influence of nonregularities of thin metal films, such as presence of granules inside the film and the roughness of surface of the film, were studied.

## Electrons in a potential energy box with an indented wall

We begin with the general case of electron inside the potential energy box. Assume rectangular potential energy box with one of the walls modified as shown in Fig. 1. Let the potential energy of the electron inside the box volume be equal to zero, and outside the box volume be equal to infinity. There is a potential energy jump from zero to infinity at any point on the walls of the box. The indents on the wall have the shape of strips having depth *a* and width *w*. Let us name the box shown on Fig. 1 Indented Potential Energy Box (IPEB) to distinguish it from the ordinary Potential Energy Box (PEB) having plane walls.

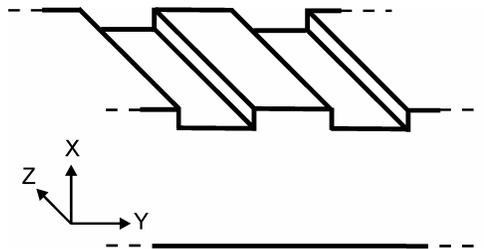

Fig. 1. 3D view of indented potential energy box.

The time independent Schrödinger equation for electron wave function inside the PEB has the form:

$$\nabla^2 \Psi + (2m/\hbar^2) E\Psi = 0 \qquad (1)$$

Here $\Psi$ is the wave function of the electron, m is the mass of the electron, and E is the energy of the electron. Let us rewrite Eq. (1) in the form of Helmholtz equation

$$(\nabla^2 + k^2)\Psi = 0 \qquad (2)$$

Where k is wave vector $k = \sqrt{2mE}/\hbar$.

Once the indent depth *a* in our particular case is supposed to be much less than thickness of the metallic film $a \ll L_x$ we can use perturbation method of solving of Helmholtz equation[2]. The idea is following: The whole volume is divided in two parts such as main volume and additional volume. It is supposed that main volume is much larger than additional volume and it defines the form of solutions for the whole composite volume. Next, solutions of the composite volume, is searched in the form of solutions of the main volume. The method is especially effective in the case main volume has simple geometry, for example rectangular geometry, allowing separation of the variables. In our case the whole volume Fig. 1 can be divided in two, Fig. 2. We regard big rectangular box as main volume and the total volume of strips as additional volume. Main volume has dimensions $L_x$, $L_y$, $L_z$. The solutions of Eq. (2) for such volume are well known. Because rectangular shape, solutions are found using method of separating of the variables. Solutions are plane waves having discrete spectrum.

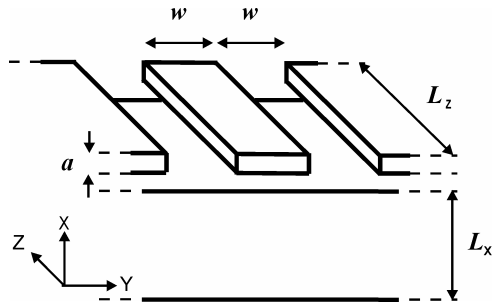

Fig.2 Potential energy box with indented wall divided it two volumes. Main volume has dimensions $L_x, L_y, L_z$, and one of the strips of additional volume has dimensions *a, w, $L_z$*.

$$k^{mx}{}_n = \pi n/L_x, \quad k^{my}{}_j = \pi j/L_y, \quad k^{mz}{}_i = \pi i/L_z \tag{3}$$

Here $k^{mx}$, $k^{my}$, $k^{mz}$ are x, y, z components of wave vectors of main volume and n, m, i=1, 2, 3,… In the same manner spectrum of the single strip of the additional volume is following:

$$k^{ax}{}_p = \pi p/a \quad k^{ay}{}_q = \pi q/w, \quad k^{az}{}_i = \pi i/L_z. \tag{4}$$

Here $k^{xp}$, $k^{yq}$ and $k^{zi}$ are components of wave vectors of one strip of additional volume, p,q=1, 2, 3… and *a* and *w* are dimensions of strip as shown on Fig. 2. Suppose main volume has length of $L_y$ in Y direction. Let us assume $L_y = l \cdot 2w$, where *l* is integer. Once solutions of main volume are, periodic in Y direction, functions and additional volume contains periodic in Y direction strips, we can find solutions of composite volume in the way of matching single strip to the main volume. Let $\Psi_m(x, y, z)$ be wave function of electron in main volume and $\Psi_a(x, y, z)$ be wave function in additional volume. The matching conditions will be $\Psi_m = \Psi_a$ and equation of partial derivatives from two sides, for all points of connection area. In main volume $\Psi_m = 0$ for all points of the walls. In additional volume $\Psi_a = 0$ for all points of the walls. Obviously $\Psi_m = \Psi_a$, for all points of connecting area is satisfied automatically. Equations of partial derivatives $\partial \Psi_m/\partial x = \partial \Psi_a/\partial x$, $\partial \Psi_m/\partial y = \partial \Psi_a/\partial y$, $\partial \Psi_m/\partial z = \partial \Psi_a/\partial z$ leads to equation of wave vector components in two volumes $k^{mx} = k^{ax}$, $k^{my} = k^{ay}$, $k^{mz} = k^{az}$. Two volumes on Fig. 2 have same spectrums along Z axis (Eq.(3), Eq.(4)). Obviously the matching of two volumes happens automatically in Z direction $k^{mz}{}_i = k^{az}{}_i = \pi i/L_z$. Along X direction we will have to match two discrete spectrums ($k^{mx}{}_n = \pi n/L_x$ and $k^{ax}{}_p = \pi p/a$)

$$k^{cx}{}_{np} = k^{mx}{}_n \cap k^{ax}{}_p = (\pi n/L_x) \cap (\pi p/a) \tag{5}$$

Here $k^{cx}{}_{np}$ is the spectrum of the composite volume in X direction. Eq. (5) means that during matching we have to select wave vectors from the spectrum of main volume, having such n that, when multiplied by $a/L_x$ it returns natural number q. Matching condition could be written as n $(a/L_x) \in N$. To obtain analytical result let us find maximum density of solutions for composite volume in X direction. According to Eq. (5)

the spectrum for composite volume is intersection of spectrum of main volume and spectrum of additional volume. Because $L_x > a$ spectrum of main volume is more dense than spectrum of additional volume. Intersection of large solution set (spectrum of main volume) and small solution set (spectrum of additional volume) is maximum when small set is subset of large set. We can write: $(\pi q/a) \in (\pi n/L_x)$ and $(L_x/a)q \in n$. Last can happen only when $L_x/a = o$, where $o$ is natural number. Once we maximized the number of solutions we can write

$$k^{cx}_{np} = \pi p/a \tag{6}$$

and use Eq. (6) for further calculations keeping in the mind that we are calculating case of maximum solution set (or spectrum density).

Next we will match solutions of two rectangular volumes along Y direction in the similar way.

$$k^{cy}_{jq} = k^{my}_{j} \cap k^{ay}_{q} = (\pi j/L_y) \cap (\pi q/w). \tag{7}$$

Here $k^{cy}_{jq}$ is spectrum of composite volume in Y direction. We maximize solutions like we did it for X direction and find:

$$k^{cy}_{jq} = \pi q/w. \tag{8}$$

Finally we have following spectrum for composite volume:

$$k^{cx}_{np} = \pi p/a,\ k^{cy}_{jq} = \pi q/w,\ k^{cz}_{ii} = \pi i/L_z. \tag{9}$$

Let us rewrite spectrums for PEB and IPEB.

$$k^{x}_{n} = \pi n/L_x,\ k^{y}_{j} = \pi j/L_y,\ k^{z}_{i} = \pi i/L_z \qquad \text{for PEB} \tag{10a}$$

$$k^{x}_{p} = \pi p/a,\ k^{y}_{q} = \pi q/w,\ k^{z}_{i} = \pi i/L_z \qquad \text{for IPEB} \tag{10b}$$

In last formulas we skip some working indexes used in this section to simplify further presentation.

Formulas 10 are obtained using perturbation method of solving of Helmholtz equation. This method assumes that additional volume is much less than main volume $(a/2L_x) \ll 1$. Special attention should be paid to the limit of very low $a$ and $w$. Case $a, w \to 0$ has following physical interpretation. Standing waves in main volume ignore additional volume because of wave diffraction on it. If we assume that wave starts ignoring nonregularityes with dimensions less than it's wavelength, we will have to make following corrections in formulas 10b. It is valid for $k^x{}_p > 2\pi/a$ and $k^y{}_q > 2\pi/w$ or p,q=2, 3, 4…. For the range $0 < k^x < 2\pi/a$ and $0 < k^y < 2\pi/w$ Eq. 10a should be used for IPEB instead of Eq. 10b. In practice dimensions *a, w* are such that only first few k should be added to Eq. 10(b).

For the case of large $a/2L_x$ other methods were used. General solution of Eq. (2) in such a complicated geometry exhibits several problems. Complicated surface shape, does not allow finding of orthogonal coordinate system which will allow separating of variables. Because of it boundary conditions may be written only in the form of piecewise regular functions. General solution of Eq. (2) usually contains infinite sums. However there are methods[2] which allow obtaining of dispersion equation and calculation of wave vector. Helmholtz equation is frequently used for calculating electromagnetic field in electromagnetic resonator cavities, waveguides and delay lines. We found that there are following analogies between electron wave function inside the metal of our geometry and electromagnetic filed inside the electromagnetic delay lines. First our geometry matches geometry of corrugated waveguide delay line[3]. Second, the same Eq. (2) is used for description in both cases. Third, boundary condition for electromagnetic wave inside the corrugated waveguide $\varepsilon=0$ (here $\varepsilon$ is electric component of electromagnetic wave) for walls of conductive waveguide, exactly matches boundary condition for electron, $\Psi=0$ outside the metal. Fourth, any wave can be presented as sum of plane waves in both cases. We also found that analogies of that kind are described in the literature[4]. Because of it, for the case of high *a*, we utilize method of solution of Helmholtz equation inside the corrugated waveguides [2,3]. Method is based on solving of transcendental equation. We found numerical solutions for the transcendental equation for high $a/2L_x$ and obtained same result, reduction of the spectrum density for IPEB relative to PEB.

Once for real thin films $L_z \gg L_x, w$, it is expected that our structure will not exhibit quantum features in Z direction. Therefore it is reasonable to consider a model of the electron motion in a two dimensional (2D) region delimited by line X=0 from one side, and periodic curve on the opposite side. In that case we can regard 2D Helmholtz equation and use special methods of it's solution. Particularly, we used the Boundary Integral Method (BIM) that is especially effective for the low-energetic part of the spectrum and has been widely employed for last years for studying 2D nano-systems[7].

The BIM method implies consideration of an appropriate integral equation instead of the Helmholtz equation. We have applied the corresponding numerical algorithm[5] to a finite number of periods and calculated the lowest few ten energy levels. We note that computation of higher energy levels with reasonable accuracy demands rapidly increasing machine-time. Obtained spectrum of transverse wave vectors also show reduction in spectral density.

In the end of this section it is worthwhile to note that above discussed effect of the energy spectrum reduction is closely related to the so-called quantum billiard problem. The studied 2D system represents the modification of quantum billiard. Unlike the circle and rectangular billiards the corrugated boundary makes the system under consideration non-integrable. This means that the number of degrees of freedom exceed the number of constants of motion. Such billiards constitute chaotic systems. The distinction between the spectra of chaotic and non-chaotic(regular) systems is exhibited by energy level spacing distribution [5,6]. Namely, the consecutive energy levels are likely to attract each other in the case of integrable system, while they repel each other in the case of chaotic system.

## Free electrons in low-dimensional metal with modified wall

To investigate how periodic indents change quantum states inside the low-dimensional metal we will use quantum model of free electrons. Free electrons inside the metal form a Fermi gas. Cyclic boundary conditions of Born-Carman

$$k^x_n = 2\pi n/L_x, \quad k^y_j = 2\pi j/L_y, \quad k^z_i = 2\pi i/L_z \tag{11}$$

are used instead of Eq. (3). Here n, j, i=0, ±1, ±2, ±3... The result of the theory is Fermi sphere in k space. All possible quantum states are occupied until $k_F$ at T=0. However for T>0 there are two types of free electrons inside the Fermi gas. Electrons with $k \approx k_F$ interact with their environment and define the transport properties of metals such as charge and heat transport. Electrons with $k << k_F$ do not interact with environment because all quantum states nearby are already occupied by other electrons (it becomes forbidden to exchange small amount of energy with environment). Such electrons are ballistic and have formally infinite mean free path. This feature allows us to regard them as planar waves, traveling between the walls of the metal (if the distance between walls is not too high). Further we will concentrate on such ballistic electrons. Once we work with electrons with infinite (or very long) mean free path, we can regard the low-dimensional metal as potential energy box and extrapolate calculations of previous section to it.

We start from comparing the volume of elementary cell in k space for thin metal films with and without periodic indents. From Equations (10a, 10b) and Eq. (11) we have:

$$dV_k = 8\pi^3/(L_x L_y L_z) \quad \text{and} \quad dV_{km} = 8\pi^3/(awL_z) \tag{12}$$

Here $dV_k$ is volume of elementary cell in k space for plain film and $dV_{kin}$ is volume of elementary cell in k space for indented film. For the ratio of volumes we get $(dV_k/ dV_{kin}) = (aw)/(L_x L_y)$. Ratio do not depends on $L_z$. Because each electron occupy more volume of k space in the case of indented metal some electrons will have to occupy quantum states with $k>k_F$. Consequently the Fermi wave vector and the corresponding Fermi energy level of indented metal film will increase Fig.3.

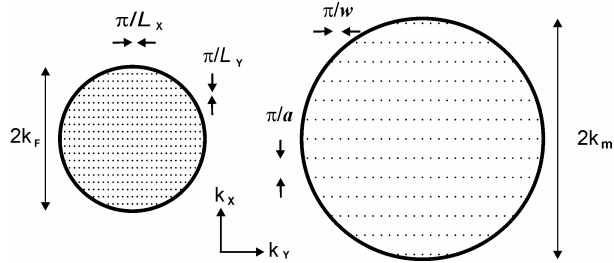

Fig.3 Section of Fermi sphere is k space for PEB (left) and IPEB (right).

Next we calculate the maximum wave vector $k_m$ at T=0 for indented metal film. Assume that metal lattice is cubic, metal is single valence and the distance between atoms is d. The volume of metal box shown on Fig.1 is

$$V = L_y L_z (L_x + a/2). \tag{13}$$

The number of atoms inside the metal of that volume is $s=V/d^3$. The number of free electrons is equal to s and we have

$$q = L_y L_z (L_x + a/2)/d^3, \tag{14}$$

for the number of free electrons. The total volume occupied by all electrons in k space will be:

$$V_m = (s/2) \, dV_{kin} = (4/3)\pi k_m^3. \tag{15}$$

Here $k_m$ is maximum possible k in the case of indented metal film and $V_m$ is the volume of the modified Fermi sphere in k space. Each **k** contains two quantum states occupied by two electrons with spins 1/2 and –1/2. That's why coefficient 2 appeared in Eq.(15). From Eq. (12), (13), (14), (15) we calculate the radius of modified Fermi sphere as

$$k_m = (1/d)[3\pi^2(L_y(L_x + a/2)/(aw)]^{1/3} . \qquad (16)$$

The radius of a Fermi sphere $k_F$ for an ordinary metal film does not depend on its dimensions and equals to $k_F = (1/d)(3\pi^2)^{1/3}$. Comparing the last with Eq. (16) we have

$$k_m = k_F [L_y(L_x + a/2)/(aw)]^{1/3} . \qquad (17)$$

Formula (17) shows the increase of the radius of the Fermi sphere in the case of low-dimensional, indented metal film in comparison with the same metal film with plane surface (Fig.3).

According to $E \sim k^2$ the Fermi energy in the low-dimensional metal film with the indented surface will relate to the Fermi energy in the same metal film with the plane surface as follows:

$$E_m = E_F [L_y(L_x + a/2)/(aw)]^{2/3} . \qquad (18)$$

If we assume $h \ll L_x$, Eq. (18) could be rewritten in the following simple form:

$$E_m = E_F (L_x L_y / aw)^{2/3} . \qquad (19)$$

As mentioned above, case of $a, w \to 0$ has different physical interpretation (wave diffraction on small volume) and should not be regarded as singularity in Eq. (19). Next question is what value of $L_y$ should be used in Eq. (19) in the case of infinite length of the structure in Y direction Fig. 1. We remember here that wave properties of electron in solids are limited by it's mean free path $\sigma$. Consequently maximum value of $L_y$ is $L_{ymax} = \sigma$. Because of it, for thin film (structure, infinite in Y direction) we can write:

$$E_m = E_F [(L_x/a)(\sigma/w)]^{2/3} . \qquad (20)$$

We remember here that calculation was made for the case Eq. (6), Eq. (8) are valid, or for the case of maximized possible quantum states.

An obvious question emerges: assume we made ratio $L_x L_y/aw$ is high enough for $E_m$ to exceed the vacuum level. What will happen? If we assume that one electron has energie greater than the vacuum level, it will leave the metal. The metal, as a result, will charge positively, and the bottom of the potential energy box will go down on the energy scale, because the metal is now charged. Once the energy level at the bottom of the potential energy box decreases, vacant place for electron will appear at the top region of the potential energy box. Electron that left the metal will return because of electrostatic force and occupy the recently released free energy level. Accordingly, $E_m$ will not exceed the vacuum level. Instead, the energy level at the bottom of the potential energy box will go down exactly at such distance to allow the potential energy box to carry all the electrons needed for electrical neutrality of the metal.

In real low-dimensional metals the surface is never ideally plane. Roughness of the surface strongly limits the increase of the Fermi level. This limits will be discussed in more detail in next section.

The dimensional quantum effects in ultra thin metal films, using rectangular potential box model and quantum model of free electrons were studied in [8].

## Problems of practical realization and possible solutions

Regarded type of structure could be obtained by depositing a thin metal film on the insulator substrate, and then etching the indents inside the metal film. What are the limitations? De Broglie wave diffraction will take place on the indents. Diffraction on the indents will lead to the wave "ignoring" the indent, which changes all the calculations above. Consequently, the results obtained are valid only when the diffraction of the wave on the indent is negligible, or $\lambda<<w$. Here $\lambda=2\pi/k_1$ is the de Broglie wavelength of the electron with wave vector $k_1$. In the case indent width is $w>\sigma$, wave properties of the electron will not propagate on many indents along Y axis. In that case wave interference will have only local character and the effect will not depend on $w$ and $L_y$. Values describing the Y dimension will not be included in Eq. (20), like values describing Z dimension are not included. This 1D case was analyzed in [9, 10].

There are some requirements to the homogeneity of the thin metal film. First, the film structure should be as close to single crystal as possible. This requirement arise because the free electron wave function should be continuous through the whole thickness of film $L_x+a$, which means that the metallic film cannot be granular. If the thin metallic film is granular, the wave function will have an interruption on the border of grains, and the indented wall's influence will be compromised. It is necessary to note here that lattice impurities do not influence free electrons with energies $E<<E_F$. In order to interact with an impurity inside the lattice, an electron should exchange energy (small amount of energy) with impurity. That type of energy exchange is forbidden because all possible quantum states nearby are already occupied. Because of this the mean free path of an electron, having energy $E \ll E_F$, is very long. Consequently the material of the film can have impurities, but the film should not be granular.

Surface roughness should be minimized, as it leads to the scattering of de Broglie waves. Scattering is considerable for the de Broglie wavelengths, less than the roughness of the surface. Substrates with a roughness of 5 Å are commercially available. Metal film deposited on such a substrate can also have a surface with the same roughness. The de Broglie wavelength of a free electron at the Fermi level is 5-10 Å in metals. Scattering of the de Broglie waves of electrons having energies $E \approx E_F$ will be considerable. Corresponding energy levels will be smoothed. Smoothing of energy levels decreases the lifetime and leads to continuous energy spectrum instead of a discrete one. Fig. 4 shows Fermi and vacuum levels of some single valence metals on the energy scale (left Y axis) and simultaneously on the scale of de Broglie wavelength (right Y axis) of the electron calculated from $\lambda = 2\pi\hbar/\sqrt{2mE}$. Fig. 4 demonstrates that 5Å roughness of the surface is enough to eliminate energy barrier for such metals as Cs and Na. The same 5Å roughness creates a gap from zero to approximately Fermi level in the energy spectrum of such metals as Au and Ag. As result, it becomes possible to reduce the work function of Cs and Na to zero, in the case of 5Å surface roughness. But the same 5Å roughness on the surface of Au or Ag, will allow a reduction of work function only by 0.5-1 eV.

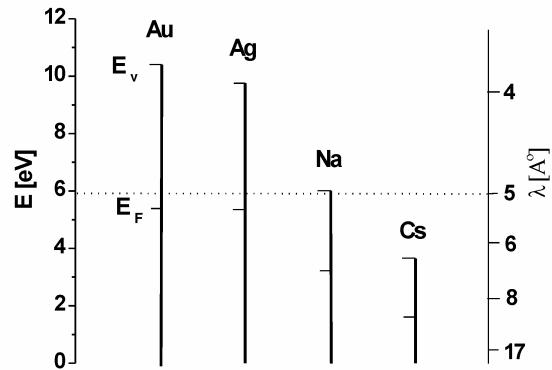

Fig. 4 Energy diagrams of some single valence metals on the scale of de Broglie wavelength.

The depth of the indent should be much more than the surface roughness. Consequently, the minimum possible $a$ is 30-50Å. According to Eq. (20) the minimum possible $w$ will be 300-500Å. The primary experimental limitation in the case of the structure shown on Fig.4 is that the ratio $(L_x/a)>5$, in order to achieve a maximum work function reduction. Consequently, the thickness of the metal film should be more than 300Å. Usually films of such thickness still repeat the substrate surface shape, and film surface roughness will not exceed the roughness of the base substrate. However, the same is not true for metal films with a thickness of 1000Å and more, because a thick film surface does not follow the surface of the substrate. That puts another limit $15 \geq (L_x/a) \geq 5$ on the dimensions of the film. Other possible structures, such as single crystal, will not be limited by the same requirements.

It should be noted that spectrum density will change if there a shift in periodicity of the indents in Y direction. In experiment it is difficult to keep exact periodicity, especially in the case of low width of the indent. In such case neighboring indents will have different spectrum $k^{ay}_q = \pi q/w$, depending on their individual width $w$. Then one more spectrum should be included in Eq. (8) and we will have intersection of three sets instead of two. It will decrease the spectrum density.

And finally there are some limits on the selection of materials which could be used for a thin film. Most metals oxidize under influence of the atmosphere. Even when placed in vacuum, metals oxidize with time, because of the influence of residual gases. Typical metal oxides have a depth of 50-100Å which is considerable on the depth scale discussed. Because of those limitations, gold is best material which can be used in experiment.

Effect was observed in thin Au and Nb films. Thin films with indented surface were fabricated and their work function reduction was observed [11]. Experimental results are in qualitative agreement with the theory. Particularly, experiments show that work function reduction strongly depends on the structure of the indented film. Amorphous films show much more reduction in work function than polycrystalline films made from the same material. Work function reduction in samples, depend on the depth of the indents, as predicted by theory. Experiments did not show quantitative agreement with the theory. One of the reasons could be that effect strongly depends on the structure of the film.

Particularly effect dramatically depends on the value of mean free path of electrons below Fermi level. In the theory we assume that mean free path of such electrons is much more than film thickness. Most probably we have not realized such condition in our resent measurements made at T=300 K. Finite roughness of the surface of Au films, cause de Broglie wave scattering and reduce the effect. Roughness of the surface is not included in calculations, and probably it is one more reason why we do not have qualitative agreement between experiment and theory.

## Conclusions

To investigate new quantum interference effect in low-dimensional metals, behavior of free electrons in the potential energy box of special geometry was studied. It was shown that, when periodic indents are introduced in the plain wall of a rectangular potential energy box, spectrum density of possible solutions of Schrödinger equation reduce dramatically. Once the number of possible quantum states decrease, electrons have to occupy higher energy levels because of Pauli exclusion principle. Results obtained for electrons in the modified potential energy box were extrapolated to the case of the free ballistic electrons inside the low-dimensional metal film. Electron distribution function of low-dimensional metal change. Fermi energy increase and consequently work function decrease. Limiting factors of the effect are metal surface roughness, and finite mean free path of ballistic electrons inside low-dimensional metal. Because surface roughness limiting factor, the magnitude of the effect will be higher for materials having low value of Fermi energy and low value of work function. Because electron mean free path limiting factor, effect will be higher in the case of single crystal or amorphous structure of the film. Recent experiments demonstrate good quantitative agreement with theory. Increase in the Fermi level and the corresponding decrease of the work function of the thin films will have practical use for devices working on the basis of electron emission and electron tunneling. In addition, such layers will be useful in the semiconductor industry, particularly for the structures in which contact potential difference between two layers plays an important role.

## Acknowledgments

The author thanks Stuart Harbron , for useful discussions. Work is financed and supported by Borealis Technical Limited, assignee of corresponding patents (US 7,074,498; 6,281,514; US 6,495,843; US 6,680,214; US 6,531,703; US 6,117,344).